\documentclass{article}
\usepackage{amsfonts}

\usepackage{graphicx}
\usepackage{amsmath}

\newtheorem{theorem}{Theorem}

\newtheorem{lemma}[theorem]{Lemma}

\newtheorem{remark}[theorem]{Remark}

\newenvironment{proof}[1][Proof]{\textbf{#1.} }{\ \rule{0.5em}{0.5em}}
\input{tcilatex}

\begin{document}

\title{Linear Quantum Feedback Networks}
\author{J. Gough$^{1}$, R. Gohm$^{1}$, M. Yanagisawa$^{2}$}
\date{}
\maketitle

\begin{abstract}
The mathematical theory of quantum feedback networks has recently been
developed \cite{QFN1} for general open quantum dynamical systems interacting
with bosonic input fields. In this article we show, for the special case of
linear dynamical systems Markovian systems with instantaneous feedback
connections, that the transfer functions can be deduced and agree with the
algebraic rules obtained in the nonlinear case. Using these rules, we derive
the the transfer functions for linear quantum systems in series, in cascade,
and in feedback arrangements mediated by beam splitter devices.
\end{abstract}

1) Institute for Mathematical and Physical Sciences, University of Wales,
Aberystwyth, Ceredigion, SY23 3BZ, Wales

2) Department of Engineering, Australian National University, Canberra, ACT
0200, Australia

\bigskip

\section{Introduction}

The aim of this paper is to deduce the algebraic rules for determining the
dynamical charactersitics of a prescribed network consisting of specified
quantum oscillator systems connected by input-output fields \cite{Gardiner}, 
\cite{Wiseman}. Physical models included cavity systems or local quantum
oscillators with a quantum optical field. The resulting dynamics is linear,
and the analysis is carried out using transfer function techniques \cite{YK1}%
, \cite{YK2}. The rules have been recently deduced in \cite{QFN1} in the
general setting for nonlinear quantum dynamical systems by first
constructing a network Hamiltonian and transfering to the interaction
picture with respect to the free flow of the fields around the network
channels. However it is of interest to restrict to linear systems for two
main reasons. Firstly, the derivation here for linear systems procedes by an
alternative method to the general nonlinear case, and we are able to confirm
the restriction of the nonlinear formula to linear systems yields the same
result. Secondly, linear systems are the most widely studied models in both
classical and quantum dynamical systems theory and so it is natural to
develop these further. There has been recent interest in the development of
coherent, or fully quantum control for linear systems \cite{GJ Series}-\cite
{NJP}\ and this paper contributes by establishing the algebraic rules for
building networks of such devices.

\section{Linear Quantum Markov Models}

The dynamical evolution of a quantum system is determined by a family of
unitaries $\left\{ V\left( t,s\right) :t\geq s\right\} $ satisfying the
propagation law $V\left( t_{3},t_{2}\right) V\left( t_{2},t_{1}\right)
=V\left( t_{3},t_{1}\right) $ where $t_{3}\geq t_{2}\geq t_{1}$. The
evolution of a state from time $s$ to a later time $t$ being then given by $%
\psi \left( t\right) =V\left( t,s\right) \psi \left( s\right) $. In a Markov
model we factor the underlying Hilbert space as $\frak{h}\otimes \mathcal{E}$
representing the system and its environment respectively and the unitary $%
V\left( t,s\right) $ couples the system specifically with the degrees of
freedom of the environment acting between times $s$ and $t$. For a bosonic
environment, we introduce input processes $b_{i}\left( t\right) $ for $%
i=1,\cdots ,n$ with the canonical commutation relations, \cite{Gardiner}, 
\begin{equation}
\left[ b_{i}\left( t\right) ,b_{j}^{\dag }\left( s\right) \right] =\delta
_{ij}\,\delta \left( t-s\right) .
\end{equation}
It is convenient to assemble these into the following column vectors of
length $n$ 
\begin{equation}
\mathbf{b}^{\text{in}}\left( t\right) =\left( 
\begin{array}{c}
b_{1}\left( 1\right) \\ 
\vdots \\ 
b_{n}\left( t\right)
\end{array}
\right) .
\end{equation}
A Markov evolution can be described equivalently by the
chronological-ordered and Wick-ordered expressions 
\begin{equation*}
V\left( t,s\right) =\;\vec{T}\exp -i\int_{s}^{t}\Upsilon \left( \tau \right)
d\tau \equiv \;:\exp -i\int_{s}^{t}\Upsilon _{\text{Wick}}\left( \tau
\right) d\tau :
\end{equation*}
where the stochastic Hamiltonian is (with $E_{ij}^{\dag }=E_{ji}$ and $%
K^{\dag }=K$) 
\begin{equation*}
\Upsilon \left( t\right) =\sum_{i,j=1}^{n}E_{ij}\otimes b_{i}^{\dag }\left(
t\right) b_{j}\left( t\right) +\sum_{i=1}^{n}F_{i}\otimes b_{i}^{\dag
}\left( t\right) +\sum_{j=1}^{n}F_{j}^{\dag }\otimes b_{j}\left( t\right)
+K\otimes 1,
\end{equation*}
and the Wick-ordered generator is given by \cite{G Wong-Zakai} 
\begin{eqnarray*}
-i\Upsilon _{\text{Wick}}\left( t\right) &=&\sum_{i,j=1}^{n}(S_{ij}-\delta
_{ij})\otimes b_{i}^{\dag }\left( t\right) b_{j}\left( t\right)
+\sum_{i=1}^{n}L_{i}\otimes b_{i}^{\dag }\left( t\right) \\
&&-\sum_{j=1}^{n}L_{i}^{\dag }S_{ij}\otimes b_{j}\left( t\right) -\left( 
\frac{1}{2}\sum_{i=1}^{n}L_{i}^{\dag }L_{i}-iH\right) \otimes 1.
\end{eqnarray*}
The Wick-ordered coefficients are given by the Stratonovich-Ito conversion
formulae, see appendix, 
\begin{equation}
S=\frac{1-\frac{i}{2}E}{1+\frac{i}{2}E},\quad L=-i\frac{1}{1+\frac{i}{2}E}%
F,\quad H=E_{00}+\frac{1}{2}\func{Im}F\frac{1}{1+\frac{i}{2}E}F^{\dag }.
\label{Strat-Ito}
\end{equation}
Note that $H$ is selfadjoint, and that $S$ is a unitary matrix whose entries
are operators on $\frak{h}$:$\sum_{k=1}^{n}S_{ik}S_{jk}^{\dag }=\delta
_{ij}=\sum_{k=1}^{n}S_{ki}^{\dag }S_{kj}$. In fact, we may write $S=e^{-iJ}$
with $J=2\arctan \dfrac{E}{2}$.

In differential form we have 
\begin{eqnarray*}
\frac{d}{dt}V\left( t,s\right) &=&\;-i:\Upsilon _{\text{Wick}}\left(
t\right) V\left( t,s\right) : \\
&\equiv &\sum_{i,j=1}^{n}b_{i}^{\dag }\left( t\right) (S_{ij}-\delta
_{ij})V\left( t,s\right) b_{j}\left( t\right) +\sum_{i=1}^{n}b_{i}^{\dag
}\left( t\right) L_{i}V\left( t,s\right) \\
&&-\sum_{j=1}^{n}L_{i}^{\dag }S_{ij}V\left( t,s\right) b_{j}\left( t\right)
-\left( \frac{1}{2}\sum_{i=1}^{n}L_{i}^{\dag }L_{i}-iH\right) V\left(
t,s\right) .
\end{eqnarray*}
Note that all the creators appear on the left and all annihilators on the
right. This equation can be interpreted as a quantum stochastic differential
equation \cite{Gardiner}, \cite{HP}, \cite{partha}.

We sketch the system plus field as a two port device having an input and an
output port.

\begin{center}
%TCIMACRO{
%\TeXButton{figure 1}{\setlength{\unitlength}{.04cm}
%\begin{picture}(120,45)
%\label{pic1}
%\thicklines
%\put(45,10){\line(0,1){20}}
%\put(45,10){\line(1,0){30}}
%\put(75,10){\line(0,1){20}}
%\put(45,30){\line(1,0){30}}
%\thinlines
%\put(48,20){\vector(-1,0){45}}
%\put(120,20){\vector(-1,0){20}}
%\put(120,20){\line(-1,0){48}}
%\put(50,20){\circle{4}}
%\put(70,20){\circle{4}}
%\put(100,26){input, ${\bf b}^{\rm in}$}
%\put(48,35){system}
%\put(-10,26){output, ${\bf b}^{\rm out}$}
%\end{picture}
%}}%
%BeginExpansion
\setlength{\unitlength}{.04cm}
\begin{picture}(120,45)
\label{pic1}
\thicklines
\put(45,10){\line(0,1){20}}
\put(45,10){\line(1,0){30}}
\put(75,10){\line(0,1){20}}
\put(45,30){\line(1,0){30}}
\thinlines
\put(48,20){\vector(-1,0){45}}
\put(120,20){\vector(-1,0){20}}
\put(120,20){\line(-1,0){48}}
\put(50,20){\circle{4}}
\put(70,20){\circle{4}}
\put(100,26){input, ${\bf b}^{\rm in}$}
\put(48,35){system}
\put(-10,26){output, ${\bf b}^{\rm out}$}
\end{picture}
%
%EndExpansion

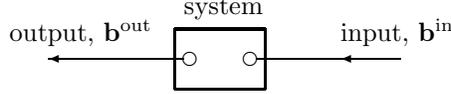
Figure 1: input-output component
\end{center}

The output fields are defined by $b_{i}^{\text{out}}\left( t\right) =V\left(
t,0\right) ^{\dag }b_{i}\left( t\right) V\left( t,0\right) $ and we have the
input-output relation 
\begin{equation*}
b_{i}^{\text{out}}\left( t\right) =\sum_{j=1}^{n}S_{ij}\left( t\right)
b_{j}\left( t\right) +L_{i}\left( t\right) ,
\end{equation*}
where $S_{ij}\left( t\right) =V\left( t,0\right) ^{\dag }S_{ij}V\left(
t,0\right) $ and $L_{i}\left( t\right) =V\left( t,0\right) ^{\dag
}L_{i}V\left( t,0\right) $. More compactly, $\mathbf{b}^{\text{out}}\left(
t\right) =S\left( t\right) \mathbf{b}^{\text{in}}\left( t\right) +L\left(
t\right) $.

Let $X$ be a fixed operator of the system and set $X\left( t,t_{0}\right)
=V\left( t,t_{0}\right) ^{\dag }XV\left( t,t_{0}\right) $, then we obtain
the Heisenberg-Langevin equation 
\begin{eqnarray*}
\frac{d}{dt}X\left( t,t_{0}\right) &=&V\left( t,t_{0}\right) ^{\dag }\frac{1%
}{i}[X,\Upsilon \left( t\right) ]V\left( t,t_{0}\right) \\
&=&b_{i}^{\dag }\left( t\right) V\left( t,t_{0}\right) ^{\dag }\left(
S_{ki}^{\dag }XS_{kj}-\delta _{ij}X\right) V\left( t,t_{0}\right)
b_{j}\left( t\right) \\
&&+b_{i}^{\dag }\left( t\right) V\left( t,t_{0}\right) ^{\dag }S_{ki}^{\dag
} \left[ X,L_{k}\right] V\left( t,t_{0}\right) \\
&&+V\left( t,t_{0}\right) ^{\dag }[L_{i}^{\dag },X]S_{ij}V\left(
t,t_{0}\right) b_{j}\left( t\right) \\
&&+V\left( t,t_{0}\right) ^{\dag }\left\{ \frac{1}{2}L_{k}^{\dag }\left[
X,L_{k}\right] +\frac{1}{2}[L_{k}^{\dag },X]L_{k}-i\left[ X,H\right]
\right\} V\left( t,t_{0}\right) .
\end{eqnarray*}
Note that the final term does not involve the input noises, and that the
expression in braces is a Lindbladian. In the special case where $S=1$, this
equation reduces to the class of Heisenberg-Langevin equations introduced by
Gardiner \cite{Gardiner}.

\subsection{Linear Models}

We consider a quantum mechanical system consisting of a family of harmonic
oscillators $\left\{ a_{j}:j=1,\cdots ,m\right\} $ with canonical
commutation relations $\left[ a_{j},a_{k}\right] =0=\left[ a_{j}^{\dag
},a_{k}^{\dag }\right] $ and $\left[ a_{j},a_{k}^{\dag }\right] =\delta
_{jk} $. We collect into column vectors: 
\begin{equation}
\mathbf{a}=\left( 
\begin{array}{c}
a_{1} \\ 
\vdots \\ 
a_{m}
\end{array}
\right) .
\end{equation}

Our interest is in the general linear open dynamical system and here we make
several simplifying assumptions:

\begin{itemize}
\item[1)]  The $S_{jk}$ are scalars.

\item[2)]  The $L_{j}^{\prime }s$ are linear, i.e., there exist constants $%
c_{jk}$ such that $L_{j}\equiv \sum_{k}c_{jk}a_{k}$.

\item[3)]  $H$ is quadratic, i.e., there exist constants $\omega _{jk}$ such
that $H=\sum_{jk}a_{j}^{\dag }\omega _{jk}a_{k}$.
\end{itemize}

The complex damping is $\frac{1}{2}L^{\dag }L+iH=-\mathbf{a}^{\dag }A\mathbf{%
a}$ where $A=-\frac{1}{2}C^{\dag }C-i\Omega $ with $C=\left( c_{jk}\right) $
and $\Omega =\left( \omega _{jk}\right) $. The Heisenberg-Langevin equations
for $\mathbf{a}\left( t\right) =V\left( t,0\right) \mathbf{a}V\left(
t,0\right) $ and input-output relations then simplify down to 
\begin{eqnarray}
\mathbf{\dot{a}}\left( t\right) &=&A\mathbf{a}\left( t\right) -C^{\dag }S%
\mathbf{b}(t), \\
\mathbf{b}^{\text{out}}\left( t\right) &=&S\mathbf{b}\left( t\right) +C%
\mathbf{a}\left( t\right) .
\end{eqnarray}
These linear equations are amenable to Laplace transform techniques \cite
{YK1},\cite{YK2}. We define for $\func{Re}s>0$ 
\begin{equation}
\hat{C}\left( s\right) =\int_{0}^{\infty }e^{-st}C\left( t\right) dt,
\end{equation}
where $C$ is now any of our stochastic processes. Note that $\widehat{%
\mathbf{\dot{a}}}\left( s\right) =s\mathbf{\hat{a}}\left( s\right) -\mathbf{a%
}$. We find that 
\begin{eqnarray*}
\mathbf{\hat{a}}\left( s\right) &=&-\left( sI_{m}-A\right) ^{-1}C^{\dag }S%
\mathbf{\hat{b}}^{\text{in}}\left( s\right) +\left( sI_{m}-A\right) ^{-1}%
\mathbf{a}, \\
\mathbf{\hat{b}}^{\text{out}}\left( s\right) &=&S\mathbf{\hat{b}}^{\text{in}%
}\left( s\right) +C\mathbf{\hat{a}}\left( s\right) .
\end{eqnarray*}

The operator $\mathbf{\hat{a}}\left( s\right) $ can be eliminated entirely
to give 
\begin{equation}
\mathbf{\hat{b}}^{\text{out}}\left( s\right) =\Xi \left( s\right) \mathbf{%
\hat{b}}^{\text{in}}\left( s\right) +\xi \left( s\right) \mathbf{a}
\end{equation}
where the \textit{transfer matrix function} is 
\begin{equation}
\Xi \left( s\right) =S-C\left( sI_{m}-A\right) ^{-1}C^{\dag }S
\end{equation}
and $\xi \left( s\right) =C\left( sI_{m}-A\right) ^{-1}$.

\bigskip

As an example, consider a single mode cavity coupling to the input field via 
$L=\sqrt{\gamma }a,$ and with Hamiltonian $H=\omega a^{\dag }a$. This
implies $K=\frac{\gamma }{2}+i\omega $ and $C=\sqrt{\gamma }$. If the output
picks up an additional phase $S=e^{i\phi }$, the corresponding transfer
function is then computed to be 
\begin{equation}
\Xi _{cavity}\left( s\right) =e^{i\phi }\,\frac{s+i\omega -\frac{\gamma }{2}%
}{s+i\omega +\frac{\gamma }{2}}.
\end{equation}

\subsection{The Transfer Matrix Function}

The models we consider are therefore determined completely by the matrices $%
\left( S,C,\Omega \right) $ with $S\in \mathbb{C}^{n\times n},C\in \mathbb{C}%
^{n\times m}$ and $\Omega \in \mathbb{C}^{m\times m}$. We shall use the
convention $\left[ 
\begin{tabular}{l|l}
$A$ & $B$ \\ \hline
$C$ & $D$%
\end{tabular}
\right] \left( s\right) =D+C\left( s-A\right) ^{-1}B$ for matrices $A\in 
\mathbb{C}^{m\times m},B\in \mathbb{C}^{m\times n},C\in \mathbb{C}^{n\times
m}$ and $D\in \mathbb{C}^{n\times n}$, and write the transfer matrix
function as

\begin{equation}
\Xi \left( s\right) =\left[ 
\begin{tabular}{r|r}
$A$ & $-C^{\dag }S$ \\ \hline
$C$ & $S$%
\end{tabular}
\right] \left( s\right) ,  \label{TF}
\end{equation}
where $A=-\frac{1}{2}C^{\dag }C-i\Omega $. We note the decomposition 
\begin{equation*}
\Xi =\left[ I_{n}-C\left( sI_{m}-A\right) ^{-1}C^{\dag }\right] S\equiv 
\left[ 
\begin{tabular}{r|r}
$A$ & $-C^{\dag }$ \\ \hline
$C$ & $I_{n}$%
\end{tabular}
\right] S.
\end{equation*}
In the simplest case of a single cavity mode we have 
\begin{equation*}
\Xi _{cavity}\left( s\right) =\left[ 
\begin{tabular}{r|r}
$-\frac{\gamma }{2}-i\omega $ & $-\sqrt{\gamma }e^{i\phi }$ \\ \hline
$\sqrt{\gamma }$ & $e^{i\phi }$%
\end{tabular}
\right] \left( s\right) .
\end{equation*}

\begin{lemma}
For each $\omega \in \mathbb{R}$, the transfer function $\Xi \left( i\omega
\right) \equiv \Xi \left( 0^{+}+i\omega \right) $ is unitary whenever it
exists.
\end{lemma}

\begin{proof}
The decomposition follows immediately from $\left( \ref{TF}\right) $. We
have then for instance 
\begin{multline*}
\Xi \left( 0^{+}+i\omega \right) \Xi \left( 0^{+}+i\omega \right) ^{\dag }= 
\left[ I-C\frac{1}{\frac{1}{2}C^{\dag }C+i\Omega ^{\prime }}C^{\dag }\right] %
\left[ I-C\frac{1}{\frac{1}{2}C^{\dag }C-i\Omega ^{\prime }}C^{\dag }\right]
\\
=I-C\frac{1}{\frac{1}{2}C^{\dag }C+i\Omega ^{\prime }}\left\{ \frac{1}{2}%
C^{\dag }C+i\Omega ^{\prime }+\frac{1}{2}C^{\dag }C-i\Omega ^{\prime
}-C^{\dag }C\right\} \frac{1}{\frac{1}{2}C^{\dag }C-i\Omega ^{\prime }}%
C^{\dag },
\end{multline*}
where $\Omega ^{\prime }=\Omega +\omega $. The term in braces however
vanishes identically, leaving $\Xi \left( 0^{+}+i\omega \right) \Xi \left(
0^{+}+i\omega \right) ^{\dag }=I$. The relation $\Xi \left( 0^{+}+i\omega
\right) \Xi \left( 0^{+}+i\omega \right) ^{\dag }=I$ is similarly
established.
\end{proof}

\bigskip

Whenever appropriate, we may determine $\Xi $ from its (unitary) values on
the imaginary axis by using the Hilbert transform 
\begin{equation*}
\Xi \left( s\right) =\frac{1}{2\pi i}PV\int_{-\infty }^{\infty }\frac{\Xi
\left( i\omega \right) }{\omega +is}d\omega .
\end{equation*}

In general, the real and imaginary parts of $A$ need not commute - that is, $%
\left[ C^{\dag }C,\Omega \right] $ need to be identically zero. However,
when this does occur we recover a multi-mode version of the cavity situation.

\begin{lemma}
If $A$ is a function of $C^{\dag }C$ then 
\begin{equation*}
\Xi \left( s\right) =\frac{s+\tilde{A}^{\dag }}{s-\tilde{A}^{\dag }}S,
\end{equation*}
where $\tilde{A}$ is a function of $CC^{\dag }$ and $\Xi $ may be
analytically continued into the whole complex plane.
\end{lemma}

\begin{proof}
Here we must have $A=-\frac{1}{2}C^{\dag }C-i\varepsilon \left( C^{\dag
}C\right) $ where $\varepsilon $ is a real valued function. We set $\tilde{A}%
=-\frac{1}{2}CC^{\dag }-i\varepsilon \left( CC^{\dag }\right) $. From the
identity $Cf\left( C^{\dag }C\right) C^{\dag }=CC^{\dag }f\left( CC^{\dag
}\right) $ for suitable analytic functions $f$, we have 
\begin{equation*}
\left( s-\tilde{A}\right) \Xi \left( s\right) =\left( s-\tilde{A}\right) %
\left[ I-\frac{1}{s-\tilde{A}}CC^{\dag }\right] S=\left( s-\tilde{A}%
-CC^{\dag }\right) S
\end{equation*}
however, $-\tilde{A}-CC^{\dag }=\tilde{A}^{\dag }$, and this gives the
result.

The hermitean matrices $C^{\dag }C$ and $CC^{\dag }$ will have the same set
of eigenvalues: to see this, suppose that $\phi $ is a non-zero unit
eigenvector of $CC^{\dag }$ with eigenvalue $\gamma $, then $\psi =\gamma
^{-1/2}C^{\dag }\phi $ is a unit eigenvector of $C^{\dag }C$ with the same
eigenvalue, conversely, every eigenvector $\psi $ of $C^{\dag }C$ with
non-zero eigenvalue $\gamma $ gives rise to a nonzero eigenvector $\phi
=\gamma ^{-1/2}C\psi $ of $CC^{\dag }$.

Let $CC^{\dag }$ have the spectral form $\sum_{k}\gamma _{k}E_{k}$ with real
eigenvalues $\gamma _{k}$ and corresponding eigenprojectors $E_{k}$, then we
have 
\begin{equation*}
\Xi \left( s\right) =\sum_{k}\frac{s-\frac{1}{2}\gamma _{k}+i\varepsilon _{k}%
}{s+\frac{1}{2}\gamma _{k}+i\varepsilon _{k}}E_{k}S,
\end{equation*}
where $\varepsilon _{k}=\varepsilon \left( \gamma _{k}\right) $. In
particular, the rational fraction is of modulus unity for imaginary $s\left(
=i\omega \right) $ and we may write 
\begin{equation*}
\Xi \left( 0^{+}+i\omega \right) =\sum_{k}e^{i\phi _{k}\left( \omega \right)
}E_{k}S
\end{equation*}
where $\phi _{k}\left( \omega \right) =\arg \frac{i\left( \omega
+\varepsilon _{k}\right) -\gamma _{k}/2}{i\left( \omega +\varepsilon
_{k}\right) +\gamma _{k}/2}$. Note that $\Xi \left( 0^{+}+i\omega \right) $
is clearly unitary and the limit $\omega \rightarrow 0$ is well-defined.
This limit will equal $-S$ in the special case that $K$ is selfadjoint
(i.e., $\varepsilon \equiv 0$). $\Xi $ may be analytically continued into
the negative-real part of the complex plane. The poles of $\Xi $ then form
the resolvent set of $\tilde{K}$, and the zeroes being the complex
conjugates.
\end{proof}

\bigskip

\section{Introducing Connections}

The situation depicted in the figure below is one where (some of) the output
channels are fed back into the system as an input. Prior to the connection
between output port(s) $s_{\mathsf{i}}$ and input port(s) $r_{\mathsf{i}}$
being made, we may model the component as having the total input $\mathbf{b}%
^{\text{in}}=\left( 
\begin{array}{c}
\mathbf{b}_{\mathsf{i}}^{\text{in}} \\ 
\mathbf{b}_{\mathsf{e}}^{\text{in}}
\end{array}
\right) $ and total output $\mathbf{b}^{\text{out}}=\left( 
\begin{array}{c}
\mathbf{b}_{\mathsf{i}}^{\text{out}} \\ 
\mathbf{b}_{\mathsf{e}}^{\text{out}}
\end{array}
\right) $ where the $\mathbf{b}_{j}^{\text{in}}$ and $\mathbf{b}_{j}^{\text{%
out}}$ may be multi-dimensional noises (we in fact only require the
multiplicities to agree for $j=\mathsf{i},\mathsf{e}$ respectively).

\begin{center}
%TCIMACRO{
%\TeXButton{figure 2}{\setlength{\unitlength}{.1cm}
%\begin{picture}(80,28)
%\label{pic2}
%
%
%\thicklines
%
%\put(30,5){\line(0,1){15}}
%\put(30,5){\line(1,0){20}}
%\put(50,5){\line(0,1){15}}
%\put(30,20){\line(1,0){20}}
%\thinlines
%\put(32,10){\vector(-1,0){15}}
%\put(63,10){\vector(-1,0){15}}
%\put(25,15){\line(1,0){7}}
%\put(25,15){\line(0,1){10}}
%\put(25,25){\line(1,0){30}}
%\put(55,25){\line(0,-1){10}}
%\put(55,15){\line(-1,0){7}}
%\put(25,25){\vector(1,0){15}}
%\put(33,10){\circle{2}}
%\put(33,15){\circle{2}}
%\put(47,10){\circle{2}}
%\put(47,15){\circle{2}}
%\put(35,10){$s_{\sf e}$}
%\put(35,15){$s_{\sf i}$}
%\put(42,10){$r_{\sf e}$}
%\put(42,15){$r_{\sf i}$}
%
%
%\end{picture}%
%}}%
%BeginExpansion
\setlength{\unitlength}{.1cm}
\begin{picture}(80,28)
\label{pic2}

\thicklines

\put(30,5){\line(0,1){15}}
\put(30,5){\line(1,0){20}}
\put(50,5){\line(0,1){15}}
\put(30,20){\line(1,0){20}}
\thinlines
\put(32,10){\vector(-1,0){15}}
\put(63,10){\vector(-1,0){15}}
\put(25,15){\line(1,0){7}}
\put(25,15){\line(0,1){10}}
\put(25,25){\line(1,0){30}}
\put(55,25){\line(0,-1){10}}
\put(55,15){\line(-1,0){7}}
\put(25,25){\vector(1,0){15}}
\put(33,10){\circle{2}}
\put(33,15){\circle{2}}
\put(47,10){\circle{2}}
\put(47,15){\circle{2}}
\put(35,10){$s_{\sf e}$}
\put(35,15){$s_{\sf i}$}
\put(42,10){$r_{\sf e}$}
\put(42,15){$r_{\sf i}$}

\end{picture}%
%
%EndExpansion

figure 2: A quantum system with feedback
\end{center}

The transfer matrix function takes the general form 
\begin{equation*}
\Xi \equiv \left[ 
\begin{tabular}{l|ll}
$A$ & $-\sum_{j}C_{j}^{\dag }S_{j\mathsf{i}}$ & $-\sum_{j}C_{j}^{\dag }S_{j%
\mathsf{e}}$ \\ \hline
$C_{\mathsf{i}}$ & $S_{\mathsf{ii}}$ & $S_{\mathsf{ie}}$ \\ 
$C_{\mathsf{e}}$ & $S_{\mathsf{ei}}$ & $S_{\mathsf{ee}}$%
\end{tabular}
\right] .
\end{equation*}
When we make the connection, we impose the various constraints $b_{r_{%
\mathsf{i}}\left( k\right) }^{\text{in}}\left( t\right) =b_{s_{\mathsf{i}%
}\left( j\right) }^{\text{out}}\left( t-\tau \right) $ where output field
labelled $s_{\mathsf{i}}\left( j\right) $ is to be connected to the input
field $r_{\mathsf{i}}\left( k\right) $ where $\tau >0$ is the time delay. We
assume the idealized situation of instantaneous feedback $\tau \rightarrow
0^{+}$. To avoid having to match up the labels of the internal channels, it
is more convenient to introduce a fixed labelling and write 
\begin{equation*}
\mathbf{b}_{\mathsf{i}}^{\text{out}}\left( t^{-}\right) =\eta \mathbf{b}_{%
\mathsf{i}}^{\text{in}}\left( t\right)
\end{equation*}
where $\eta $ is the adjacency matrix: 
\begin{equation*}
\eta _{sr}=\left\{ 
\begin{array}{cc}
1, & \text{if }\left( s,r\right) \text{ is an internal channel,} \\ 
0, & \text{otherwise}
\end{array}
\right.
\end{equation*}
The model with the connections is then a reduction of the original and the
remaining external fields are the input $\mathbf{b}_{\mathsf{e}}^{\text{in}}$
and output $\mathbf{b}_{\mathsf{e}}^{\text{out}}$.

\begin{theorem}
Let $\left( \eta -S_{\mathsf{ii}}\right) $ be invertible. The feedback
system described above has input-output relation $\mathbf{\hat{b}}_{\mathsf{e%
}}^{\text{out}}=\Xi _{\mathrm{red}}\mathbf{\hat{b}}_{\mathsf{e}}^{\text{in}%
}+\xi _{\mathrm{red}}\mathbf{a}$ and the reduced transfer matrix function 
\begin{equation*}
\Xi _{\mathrm{red}}\equiv \left[ 
\begin{tabular}{r|r}
$A_{\mathrm{red}}$ & $-C_{\mathrm{red}}^{\dag }S_{\mathrm{red}}$ \\ \hline
$C_{\mathrm{red}}$ & $S_{\mathrm{red}}$%
\end{tabular}
\right] ,\quad \xi _{\mathrm{red}}\equiv C_{\mathrm{red}}\frac{1}{s-A_{%
\mathrm{red}}},
\end{equation*}
where 
\begin{eqnarray}
S_{\mathrm{red}} &=&S_{\mathsf{ee}}+S_{\mathsf{ei}}\left( \eta -S_{\mathsf{ii%
}}\right) ^{-1}S_{\mathsf{ie}},  \notag \\
C_{\mathrm{red}} &=&S_{\mathsf{ei}}\left( \eta -S_{\mathsf{ii}}\right)
^{-1}C_{\mathsf{i}}+C_{\mathsf{e}},  \notag \\
A_{\mathrm{red}} &=&A-\sum_{j=\mathsf{i},\mathsf{e}}C_{j}^{\dag }S_{j\mathsf{%
i}}\left( \eta -S_{\mathsf{ii}}\right) ^{-1}C_{\mathsf{i}}.
\end{eqnarray}
\end{theorem}

\begin{proof}
The dynamical equations can be written as 
\begin{eqnarray*}
\mathbf{\dot{a}}\left( t\right) &=&A\mathbf{a}\left( t\right)
-\sum_{j,k}C_{j}^{\dag }S_{jk}\mathbf{b}_{k}^{\text{in}}\left( t\right) , \\
\mathbf{b}_{j}^{\text{out}}\left( t\right) &=&\sum_{k=\mathsf{i},\mathsf{e}%
}S_{jk}\mathbf{b}_{k}^{\text{in}}\left( t\right) +C_{j}\mathbf{a}\left(
t\right) .
\end{eqnarray*}
Now the constraint $\eta \mathbf{b}_{\mathsf{i}}^{\text{in}}=\mathbf{b}_{%
\mathsf{i}}^{\text{out}}$ implies that 
\begin{equation*}
\mathbf{b}_{\mathsf{i}}^{\text{in}}\left( t\right) =\left( \eta -S_{\mathsf{%
ii}}\right) ^{-1}(S_{\mathsf{ie}}\mathbf{b}_{\mathsf{e}}^{\text{in}}\left(
t\right) +C_{\mathsf{i}}\mathbf{a}\left( t\right) ),
\end{equation*}
and so 
\begin{eqnarray*}
\mathbf{\dot{a}}\left( t\right) &=&[A-\sum_{j=\mathsf{i},\mathsf{e}%
}C_{j}^{\dag }S_{j\mathsf{i}}\left( \eta -S_{\mathsf{ii}}\right) ^{-1}C_{%
\mathsf{i}}]\mathbf{a}\left( t\right) \\
&&-\sum_{j=\mathsf{i},\mathsf{e}}C_{j}^{\dag }\left( S_{j\mathsf{e}}+S_{j%
\mathsf{i}}\left( \eta -S_{\mathsf{ii}}\right) ^{-1}S_{\mathsf{ie}}\right) 
\mathbf{b}_{\mathsf{e}}^{\text{in}}\left( t\right)
\end{eqnarray*}
or 
\begin{equation*}
\mathbf{\hat{a}}\left( s\right) =-\frac{1}{s-A_{\mathrm{red}}}\sum_{j=%
\mathsf{i},\mathsf{e}}C_{j}^{\dag }\left( S_{j\mathsf{e}}+S_{j\mathsf{i}%
}\left( \eta -S_{\mathsf{ii}}\right) ^{-1}S_{\mathsf{ie}}\right) \mathbf{%
\hat{b}}_{\mathsf{e}}\left( s\right) +\frac{1}{s-A_{\mathrm{red}}}\mathbf{a},
\end{equation*}
with $A_{\mathrm{red}}$ as above. Consequently, 
\begin{eqnarray*}
\mathbf{\hat{b}}_{\mathsf{e}}^{\text{out}} &=&S_{\mathsf{ei}}\mathbf{\hat{b}}%
_{\mathsf{i}}^{\text{in}}+S_{\mathsf{ee}}\mathbf{\hat{b}}_{\mathsf{e}}^{%
\text{in}}+C_{\mathsf{e}}\mathbf{\hat{a}} \\
&=&S_{\mathrm{red}}\mathbf{\hat{b}}_{\mathsf{e}}^{\text{in}}+C_{\mathrm{red}}%
\mathbf{\hat{a}} \\
&=&\Xi _{\mathrm{red}}\mathbf{\hat{b}}_{\mathsf{e}}^{\text{in}}+\xi _{%
\mathrm{red}}\mathbf{a}
\end{eqnarray*}
where 
\begin{eqnarray*}
\Xi _{\mathrm{red}} &=&S_{\mathrm{red}}-\sum_{j=\mathsf{i},\mathsf{e}}C_{%
\mathrm{red}}\frac{1}{s-A_{\mathrm{red}}}C_{j}^{\dag }\left( S_{j\mathsf{e}%
}+S_{j\mathsf{i}}\left( \eta -S_{\mathsf{ii}}\right) ^{-1}S_{\mathsf{ie}%
}\right) , \\
\xi _{\mathrm{red}} &=&C_{\mathrm{red}}\frac{1}{s-A_{\mathrm{red}}},
\end{eqnarray*}
and \ $S_{\mathrm{red}}$, $C_{\mathrm{red}}$ are as in the statement of the
theorem.

We now show that $\sum_{j=\mathsf{i},\mathsf{e}}C_{j}^{\dag }[S_{j\mathsf{e}%
}+S_{j\mathsf{i}}\left( \eta -S_{\mathsf{ii}}\right) ^{-1}S_{\mathsf{ie}%
}]=C_{\mathrm{red}}^{\dag }S_{\mathrm{red}}$. Now 
\begin{eqnarray*}
\sum_{j=\mathsf{i},\mathsf{e}}C_{j}^{\dag }[S_{j\mathsf{e}}+S_{j\mathsf{i}%
}\left( \eta -S_{\mathsf{ii}}\right) ^{-1}S_{\mathsf{ie}}] &=&C_{\mathsf{i}%
}^{\dag }[S_{\mathsf{ie}}+S_{\mathsf{ii}}\left( \eta -S_{\mathsf{ii}}\right)
^{-1}S_{\mathsf{ie}}]+C_{\mathsf{e}}^{\dag }S_{\mathrm{red}} \\
&=&C_{\mathsf{i}}^{\dag }\eta \left( \eta -S_{\mathsf{ii}}\right) ^{-1}S_{%
\mathsf{ie}}+C_{\mathsf{e}}^{\dag }S_{\mathrm{red}},
\end{eqnarray*}
while $C_{\mathrm{red}}^{\dag }S_{\mathrm{red}}=C_{\mathsf{i}}^{\dag }(\eta
^{\dag }-S_{\mathsf{ii}}^{\dag })^{-1}S_{\mathsf{ie}}^{\dag }S_{\mathrm{red}%
}+C_{\mathsf{e}}^{\dag }S_{\mathrm{red}}$. However, 
\begin{equation*}
(\eta ^{\dag }-S_{\mathsf{ii}}^{\dag })^{-1}S_{\mathsf{ie}}^{\dag }S_{%
\mathrm{red}}=(\eta ^{\dag }-S_{\mathsf{ii}}^{\dag })^{-1}S_{\mathsf{ei}%
}^{\dag }(S_{\mathsf{ee}}+S_{\mathsf{ei}}\left( \eta -S_{\mathsf{ii}}\right)
^{-1}S_{\mathsf{ie}})
\end{equation*}
and using the identities $S_{\mathsf{ii}}^{\dag }S_{\mathsf{ii}}+S_{\mathsf{%
ei}}^{\dag }S_{\mathsf{ei}}=1$, $S_{\mathsf{ii}}^{\dag }S_{\mathsf{ie}}+S_{%
\mathsf{ei}}^{\dag }S_{\mathsf{ee}}=0$, this reduces to 
\begin{eqnarray*}
(\eta ^{\dag }-S_{\mathsf{ii}}^{\dag })^{-1}S_{\mathsf{ie}}^{\dag }S_{%
\mathrm{red}} &=&(\eta ^{\dag }-S_{\mathsf{ii}}^{\dag })^{-1}\left[ -S_{%
\mathsf{ii}}^{\dag }S_{\mathsf{ie}}+(1-S_{\mathsf{ii}}^{\dag }S_{\mathsf{ii}%
})\left( \eta -S_{\mathsf{ii}}\right) ^{-1}S_{\mathsf{ie}}\right] \\
&=&(\eta ^{\dag }-S_{\mathsf{ii}}^{\dag })^{-1}\left[ -S_{\mathsf{ii}}^{\dag
}\left( \eta -S_{\mathsf{ii}}\right) +(1-S_{\mathsf{ii}}^{\dag }S_{\mathsf{ii%
}})\right] \left( \eta -S_{\mathsf{ii}}\right) ^{-1}S_{\mathsf{ie}} \\
&=&\eta \left( \eta -S_{\mathsf{ii}}\right) ^{-1}S_{\mathsf{ie}}.
\end{eqnarray*}
Therefore $\Xi _{\mathrm{red}}=S_{\mathrm{red}}-\sum_{j=\mathsf{i},\mathsf{e}%
}C_{\mathrm{red}}\dfrac{1}{s-A_{\mathrm{red}}}C_{\mathrm{red}}^{\dag }S_{%
\mathrm{red}}$, as required.

For consistency, we should check that we have $A_{\mathrm{red}}=-\frac{1}{%
\mathsf{e}}C_{\mathrm{red}}^{\dag }C_{\mathrm{red}}-i\Omega _{\mathrm{red}}$
with $\Omega _{\mathrm{red}}$ selfadjoint. Indeed, setting $A=-\frac{1}{2}C_{%
\mathsf{i}}^{\dag }C_{\mathsf{i}}-\frac{1}{2}C_{\mathsf{e}}^{\dag }C_{%
\mathsf{e}}-i\Omega $ and substituting in for $C_{\mathrm{red}}$ and $K_{%
\mathrm{red}}$ we find after some algebra that 
\begin{equation*}
\Omega _{\mathrm{red}}=\Omega +\func{Im}\left\{ C_{\mathsf{i}}^{\dag }S_{%
\mathsf{ii}}\left( \eta -S_{\mathsf{ii}}\right) ^{-1}C_{\mathsf{i}}\right\} +%
\func{Im}\left\{ C_{\mathsf{e}}^{\dag }S_{\mathsf{ei}}\left( \eta -S_{%
\mathsf{ii}}\right) ^{-1}C_{\mathsf{i}}\right\} .
\end{equation*}
The manipulation for this is trivial except for the calculation of the term
of the form $\frac{1}{2}C_{\mathsf{i}}^{\dag }XC_{\mathsf{i}}$ where 
\begin{eqnarray*}
X &=&1+2S_{\mathsf{ii}}\left( \eta -S_{\mathsf{ii}}\right) ^{-1}-(\eta
^{\dag }-S_{\mathsf{ii}}^{\dag })^{-1}S_{\mathsf{ei}}^{\dag }S_{\mathsf{ei}%
}\left( \eta -S_{\mathsf{ii}}\right) ^{-1} \\
&\equiv &(1-\eta S_{\mathsf{ii}}^{\dag })^{-1}\left[ S_{\mathsf{ii}}\eta
^{\dag }-\eta S_{\mathsf{ii}}^{\dag }\right] \left( 1-S_{\mathsf{ii}}\eta
^{\dag }\right) ^{-1} \\
&=&2i\func{Im}\frac{S_{\mathsf{ii}}\eta ^{\dag }}{1-S_{\mathsf{ii}}\eta
^{\dag }}=2i\func{Im}\left\{ S_{\mathsf{ii}}\left( \eta -S_{\mathsf{ii}%
}\right) ^{-1}\right\}
\end{eqnarray*}
where again we use the identity $S_{\mathsf{ii}}^{\dag }S_{\mathsf{ii}}+S_{%
\mathsf{ei}}^{\dag }S_{\mathsf{ei}}=1$.
\end{proof}

\bigskip

In terms of the parameters $\left( S,L,H\right) $ with $S=\left( 
\begin{array}{cc}
S_{\mathsf{ii}} & S_{\mathsf{ie}} \\ 
S_{\mathsf{ei}} & S_{\mathsf{ee}}
\end{array}
\right) $, $L=\left( 
\begin{array}{c}
L_{\mathsf{i}} \\ 
L_{\mathsf{e}}
\end{array}
\right) =\left( 
\begin{array}{c}
C_{\mathsf{i}}\mathbf{a} \\ 
C_{\mathsf{e}}\mathbf{a}
\end{array}
\right) $ and $H=\mathbf{a}^{\dag }\Omega \mathbf{a}$, we have that the
feedback system is described by the reduced parameters $\left( S_{\mathrm{red%
}},L_{\mathrm{red}},H_{\mathrm{red}}\right) $ where 
\begin{eqnarray}
S_{\mathrm{red}} &=&S_{\mathsf{ee}}+S_{\mathsf{ei}}\left( \eta -S_{\mathsf{ii%
}}\right) ^{-1}S_{\mathsf{ie}}  \notag \\
L_{\mathrm{red}} &=&S_{\mathsf{ei}}\left( \eta -S_{\mathsf{ii}}\right)
^{-1}L_{\mathsf{i}}+L_{\mathsf{e}},  \notag \\
H_{\mathrm{red}} &=&H+\func{Im}\left\{ L_{\mathsf{i}}^{\dag }S_{\mathsf{ii}%
}\left( \eta -S_{\mathsf{ii}}\right) ^{-1}L_{\mathsf{i}}\right\} +\func{Im}%
\left\{ L_{\mathsf{e}}^{\dag }S_{\mathsf{ei}}\left( \eta -S_{\mathsf{ii}%
}\right) ^{-1}L_{\mathsf{i}}\right\} .  \notag \\
&&  \label{general feedback law}
\end{eqnarray}
The same equations have been deduced in the nonlinear case by different
arguments \cite{QFN1}. Note the identity $\func{Im}\left\{ L_{\mathsf{i}%
}^{\dag }S_{\mathsf{ii}}\left( \eta -S_{\mathsf{ii}}\right) ^{-1}L_{\mathsf{i%
}}\right\} =\func{Im}\left\{ L_{\mathsf{i}}^{\dag }\left( \eta -S_{\mathsf{ii%
}}\right) ^{-1}L_{\mathsf{i}}\right\} $.

\bigskip

\begin{remark}
Let $U$ be a unitary operator on a fixed Hilbert space $\frak{H}=\frak{H}%
_{1}\oplus \frak{H}_{2}$ which decomposes as $U=\left( 
\begin{array}{cc}
U_{11} & U_{12} \\ 
U_{21} & U_{22}
\end{array}
\right) $. The non-commutative M\"{o}bius transform $\varphi
_{U}^{2\rightarrow 1}$ is the superoperator defined by 
\begin{equation*}
\varphi _{U}^{2\rightarrow 1}\left( X\right) =U_{11}+U_{12}\left(
1-XU_{22}\right) ^{-1}XU_{21}
\end{equation*}
defined on the domain of operators $X$ on $\frak{H}_{2}$ for which the
inverse $\left( 1-XU_{22}\right) ^{-1}$ exists. The transform $\varphi
_{U}^{2\rightarrow 1}$ maps unitaries on $\frak{H}_{2}$ in its domain to
unitaries in $\frak{H}_{1}$ \cite{Young}.
\end{remark}

\begin{remark}
In particular, $S_{\mathrm{red}}$ is unitary as it equals $\varphi _{S}^{%
\mathsf{i}\rightarrow \mathsf{e}}\left( \xi \right) $ where $\xi =\eta ^{-1}$
with $\eta $ being unitary. We may expand the geometric series to write 
\begin{equation*}
S_{\mathrm{red}}=S_{\mathsf{ee}}+S_{\mathsf{ei}}\xi S_{\mathsf{ie}}+S_{%
\mathsf{ei}}\xi S_{\mathsf{ii}}\xi S_{\mathsf{ie}}+S_{\mathsf{ei}}\xi S_{%
\mathsf{ii}}\xi S_{\mathsf{ii}}\xi S_{\mathsf{ie}}+\cdots =S_{\mathsf{ee}%
}+\sum_{n=0}^{\infty }S_{\mathsf{ei}}\xi \left( S_{\mathsf{ii}}\xi \right)
^{n}S_{\mathsf{ie}}
\end{equation*}
which shows that $S_{\mathrm{red}}$ can be built up from\ contributions from
the various paths through the network. Likewise 
\begin{eqnarray*}
L_{\mathrm{red}} &=&L_{\mathsf{e}}+\sum_{n=0}^{\infty }S_{\mathsf{ei}}\xi
\left( S_{\mathsf{ii}}\xi \right) ^{n}L_{\mathsf{i}},\quad \\
H_{\mathrm{red}} &=&H+\sum_{n=0}^{\infty }\func{Im}\left\{ L_{\mathsf{i}%
}^{\dag }\left( S_{\mathsf{ii}}\xi \right) ^{n}L_{\mathsf{i}}\right\}
+\sum_{n=0}^{\infty }\func{Im}\left\{ L_{\mathsf{e}}^{\dag }S_{\mathsf{ei}%
}\xi \left( S_{\mathsf{ii}}\xi \right) ^{n}L_{\mathsf{i}}\right\} .
\end{eqnarray*}
\end{remark}

\section{Systems in Series}

As a very special case of feedback connections we consider the situation of
systems in series. This is referred to as \textit{feedforward} in
engineering.

\begin{center}
%TCIMACRO{
%\TeXButton{figure 3}{\setlength{\unitlength}{.1cm}
%\begin{picture}(100,22)
%\label{pic3}
%
%
%\thicklines
%
%\put(30,5){\line(0,1){10}}
%\put(30,5){\line(1,0){20}}
%\put(50,5){\line(0,1){10}}
%\put(30,15){\line(1,0){20}}
%
%\put(60,5){\line(0,1){10}}
%\put(60,5){\line(1,0){20}}
%\put(80,5){\line(0,1){10}}
%\put(60,15){\line(1,0){20}}
%
%\thinlines
%\put(32,10){\vector(-1,0){15}}
%\put(62,10){\vector(-1,0){14}}
%
%\put(92,10){\vector(-1,0){14}}
%
%\put(33,10){\circle{2}}
%\put(63,10){\circle{2}}
%
%\put(47,10){\circle{2}}
%\put(77,10){\circle{2}}
%
%\put(35,11){$s_2$}
%\put(65,11){$s_1$}
%\put(42,11){$r_2$}
%\put(72,11){$r_1$}
%
%
%\end{picture}%
%}}%
%BeginExpansion
\setlength{\unitlength}{.1cm}
\begin{picture}(100,22)
\label{pic3}

\thicklines

\put(30,5){\line(0,1){10}}
\put(30,5){\line(1,0){20}}
\put(50,5){\line(0,1){10}}
\put(30,15){\line(1,0){20}}

\put(60,5){\line(0,1){10}}
\put(60,5){\line(1,0){20}}
\put(80,5){\line(0,1){10}}
\put(60,15){\line(1,0){20}}

\thinlines
\put(32,10){\vector(-1,0){15}}
\put(62,10){\vector(-1,0){14}}

\put(92,10){\vector(-1,0){14}}

\put(33,10){\circle{2}}
\put(63,10){\circle{2}}

\put(47,10){\circle{2}}
\put(77,10){\circle{2}}

\put(35,11){$s_2$}
\put(65,11){$s_1$}
\put(42,11){$r_2$}
\put(72,11){$r_1$}

\end{picture}%
%
%EndExpansion

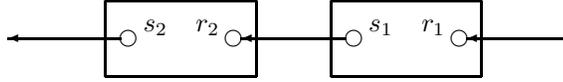
figure 3: Cascaded systems
\end{center}

The individual transfer functions before the connection $e=\left(
s_{1},r_{2}\right) $ is made are given by $\Xi _{i}=\left[ 
\begin{tabular}{r|r}
$A_{i}$ & $-C_{i}^{\dag }S_{i}$ \\ \hline
$C_{i}$ & $S_{i}$%
\end{tabular}
\right] $ with $A_{i}=-\frac{1}{2}C_{i}^{\dag }C_{i}-i\Omega _{i}$.and these
may be concatenated to give 
\begin{equation*}
\Xi =\left[ 
\begin{tabular}{l|ll}
$A_{1}+A_{2}$ & $-C_{1}^{\dag }S_{1}$ & $-C_{2}^{\dag }S_{2}$ \\ \hline
$C_{1}$ & $S_{1}$ & $0$ \\ 
$C_{2}$ & $0$ & $S_{2}$%
\end{tabular}
\right] .
\end{equation*}

To use the formula for the reduced transfer function following connection,
we must first of all identify the internal (eliminated) and external fields:
here 
\begin{equation*}
\mathbf{b}^{\text{in}}=\left( 
\begin{array}{c}
\mathbf{b}_{\mathsf{i}}^{\text{in}} \\ 
\mathbf{b}_{\mathsf{e}}^{\text{in}}
\end{array}
\right) =\left( 
\begin{array}{c}
\mathbf{b}_{2}^{\text{in}} \\ 
\mathbf{b}_{1}^{\text{in}}
\end{array}
\right) ,\quad \mathbf{b}^{\text{out}}=\left( 
\begin{array}{c}
\mathbf{b}_{\mathsf{i}}^{\text{out}} \\ 
\mathbf{b}_{\mathsf{e}}^{\text{out}}
\end{array}
\right) \equiv \left( 
\begin{array}{c}
\mathbf{b}_{1}^{\text{out}} \\ 
\mathbf{b}_{2}^{\text{out}}
\end{array}
\right) ,
\end{equation*}
and 
\begin{equation*}
\left( 
\begin{array}{cc}
S_{\mathsf{ii}} & S_{\mathsf{ie}} \\ 
S_{\mathsf{ei}} & S_{\mathsf{ee}}
\end{array}
\right) \equiv \left( 
\begin{array}{cc}
0 & S_{1} \\ 
S_{2} & 0
\end{array}
\right) ,\quad L_{\mathsf{i}}\equiv L_{1},L_{\mathsf{e}}\equiv L_{2},
\end{equation*}
with trivially $\eta =1$. The reduced transfer function is then readily
computed to be 
\begin{equation*}
\Xi _{\text{series}}=\left[ 
\begin{tabular}{r|r}
$A_{1}+A_{2}-C_{2}^{\dag }S_{2}C_{1}$ & $-\left( C_{2}^{\dag
}S_{2}+C_{1}^{\dag }\right) S_{1}$ \\ \hline
$C_{2}+S_{2}C_{1}$ & $S_{2}S_{1}$%
\end{tabular}
\right] .
\end{equation*}
Likewise we deduce the relations 
\begin{equation}
S=S_{2}S_{1},\quad L=L_{2}+S_{2}L_{1},\quad H=H_{1}+H_{2}+\func{Im}\left\{
L_{2}^{\dag }S_{2}L_{1}\right\} .  \label{special feedback law}
\end{equation}
The same equations have been deduced in the nonlinear case by different
arguments \cite{GJ Series}.

\subsection{Feedforward: Cascades}

If the two systems are truly distinct systems, that is, if they are
different sets of oscillators, then we are in the situation of properly 
\textit{cascaded} systems. In this case one would expect that the transfer
function to factor as the ordinary matrix product $\Xi _{\text{series}%
}\equiv \Xi _{2}\Xi _{1}$. We now show that this is indeed the case.

\begin{lemma}
Let $\Xi _{j}$ be transfer functions for $m_{j}$ oscillators coupled to $n$
fields $(j=1,2)$. If we consider the ampliated transfer functions for $%
m_{1}+m_{2}$ oscillators coupled to $n$ fields 
\begin{eqnarray*}
\tilde{\Xi}_{1} &=&\left[ 
\begin{tabular}{c|c}
$\left( 
\begin{array}{cc}
A_{1} & 0 \\ 
0 & 0
\end{array}
\right) $ & $\left( 
\begin{array}{c}
-C_{1}^{\dag }S_{1} \\ 
0
\end{array}
\right) $ \\ \hline
$\left( C_{1},0\right) $ & $S_{1}$%
\end{tabular}
\right] , \\
\tilde{\Xi}_{2} &=&\left[ 
\begin{tabular}{c|c}
$\left( 
\begin{array}{cc}
0 & 0 \\ 
0 & A_{2}
\end{array}
\right) $ & $\left( 
\begin{array}{c}
0 \\ 
-C_{2}^{\dag }S_{2}
\end{array}
\right) $ \\ \hline
$\left( 0,C_{2}\right) $ & $S_{2}$%
\end{tabular}
\right] ,
\end{eqnarray*}
then 
\begin{equation}
\tilde{\Xi}_{\text{series}}=\Xi _{2}\Xi _{1}.
\end{equation}
\end{lemma}

\begin{proof}
We compute this directly, 
\begin{eqnarray*}
\tilde{\Xi}_{\text{series}} &=&\left[ 
\begin{tabular}{c|c}
$\left( 
\begin{array}{cc}
A_{1} & 0 \\ 
-C_{2}^{\dag }S_{2}C_{1} & -A_{2}
\end{array}
\right) $ & $\left( 
\begin{array}{c}
-C_{1}^{\dag }S_{1} \\ 
-C_{2}^{\dag }S_{2}S_{1}
\end{array}
\right) $ \\ \hline
$\left( C_{1},C_{2}\right) $ & $S_{2}S_{1}$%
\end{tabular}
\right] \\
&=&S_{2}S_{1}+\left. 
\begin{array}{c}
(C_{1},C_{2})
\end{array}
\right. \left( 
\begin{array}{cc}
s-A_{1} & 0 \\ 
C_{2}^{\dag }S_{2}C_{1} & s-A_{2}
\end{array}
\right) ^{-1}\left( 
\begin{array}{c}
-C_{1}^{\dag }S_{1} \\ 
-C_{2}^{\dag }S_{2}S_{1}
\end{array}
\right) \\
&=&S_{2}S_{1}-\left. 
\begin{array}{c}
(C_{1},C_{2})
\end{array}
\right. \left( 
\begin{array}{ll}
\frac{1}{s-A_{1}} & 0 \\ 
-\frac{1}{s-A_{2}}C_{2}^{\dag }S_{2}C_{1}\frac{1}{s-A_{1}} & \frac{1}{s-A_{2}%
}
\end{array}
\right) \left( 
\begin{array}{c}
C_{1}^{\dag }S_{1} \\ 
C_{2}^{\dag }S_{2}S_{1}
\end{array}
\right) \\
&=&\left[ S_{2}-C_{2}\left( s-A_{2}\right) ^{-1}C_{2}^{\dag }S_{2}\right]
\times \left[ S_{1}-C_{1}\left( s-A_{1}\right) ^{-1}C_{1}^{\dag }S_{1}\right]
,
\end{eqnarray*}
giving the result.
\end{proof}

\section{Beam Splitters}

A simple beam splitter is a device performing physical superposition of two
input fields. It is described by a fixed unitary operator $T=\left( 
\begin{array}{cc}
\alpha & \beta \\ 
\mu & \nu
\end{array}
\right) \in U\left( 2\right) $: 
\begin{equation*}
\left( 
\begin{array}{c}
\mathbf{b}_{1}^{\text{out}} \\ 
\mathbf{b}_{2}^{\text{out}}
\end{array}
\right) =\left( 
\begin{array}{cc}
\alpha & \beta \\ 
\mu & \nu
\end{array}
\right) \left( 
\begin{array}{c}
\mathbf{b}_{1}^{\text{in}} \\ 
\mathbf{b}_{2}^{\text{in}}
\end{array}
\right) .
\end{equation*}
This is a canonical transformation and the output fields satisfy the same
canonical commutation relations as the inputs. The action of the beam
splitter is depicted in the figure below. On the left we have a traditional
view of the two inputs being split into two output fields. On the right we
have our view of the beam splitter as being a component with two input ports
and two output ports: we have sketched some internal detail to emphasize how
the scattering (superimposing) of inputs how ever we shall usually just draw
this as a ``black box'' component in the following.

\begin{center}
%TCIMACRO{
%\TeXButton{figure 4}{\setlength{\unitlength}{.075cm}
%\begin{picture}(100,40)
%\label{pic4}
%\thinlines
%
%\put(64,10){\vector(-1,0){9}}
%\put(64,25){\vector(-1,0){9}}
%\put(66,10){\line(1,0){13}}
%\put(66,25){\line(1,0){13}}
%\put(90,10){\vector(-1,0){9}}
%\put(90,25){\vector(-1,0){9}}
%\put(66,11){\line(1,1){13}}
%\put(66,24){\line(1,-1){13}}
%
%\put(65,10){\circle{2}}
%\put(65,25){\circle{2}}
%\put(80,10){\circle{2}}
%\put(80,25){\circle{2}}
%
%\thinlines
%\put(17,20){\vector(1,0){12}}
%\put(15,5){\vector(0,1){12}}
%\put(0,20){\vector(1,0){12}}
%\put(15,23){\vector(0,1){12}}
%\thicklines
%\put(7,12){\line(1,1){16}}
%
%\put(88,12){${\bf b}^{\rm in}_2$}
%\put(88,27){${\bf b}^{\rm in}_1$}
%\put(48,12){${\bf b}^{\rm out}_2$}
%\put(48,27){${\bf b}^{\rm out}_1$}
%
%
%\put(18,8){${\bf b}^{\rm in}_2$}
%\put(0,22){${\bf b}^{\rm in}_1$}
%\put(17,35){${\bf b}^{\rm out}_2$}
%\put(30,22){${\bf b}^{\rm out}_1$}
%
%
%\put(62,7){\dashbox(22,20)}
%
%\end{picture}
%}}%
%BeginExpansion
\setlength{\unitlength}{.075cm}
\begin{picture}(100,40)
\label{pic4}
\thinlines

\put(64,10){\vector(-1,0){9}}
\put(64,25){\vector(-1,0){9}}
\put(66,10){\line(1,0){13}}
\put(66,25){\line(1,0){13}}
\put(90,10){\vector(-1,0){9}}
\put(90,25){\vector(-1,0){9}}
\put(66,11){\line(1,1){13}}
\put(66,24){\line(1,-1){13}}

\put(65,10){\circle{2}}
\put(65,25){\circle{2}}
\put(80,10){\circle{2}}
\put(80,25){\circle{2}}

\thinlines
\put(17,20){\vector(1,0){12}}
\put(15,5){\vector(0,1){12}}
\put(0,20){\vector(1,0){12}}
\put(15,23){\vector(0,1){12}}
\thicklines
\put(7,12){\line(1,1){16}}

\put(88,12){${\bf b}^{\rm in}_2$}
\put(88,27){${\bf b}^{\rm in}_1$}
\put(48,12){${\bf b}^{\rm out}_2$}
\put(48,27){${\bf b}^{\rm out}_1$}

\put(18,8){${\bf b}^{\rm in}_2$}
\put(0,22){${\bf b}^{\rm in}_1$}
\put(17,35){${\bf b}^{\rm out}_2$}
\put(30,22){${\bf b}^{\rm out}_1$}

\put(62,7){\dashbox(22,20)}

\end{picture}
%
%EndExpansion

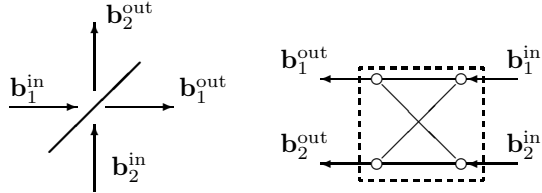
Figure 4: Beam-splitter component.
\end{center}

To emphasize that the beam splitter is an input-output device of exactly the
for we have been considering up to now, let us state that its transfer
matrix function is 
\begin{equation*}
\Xi _{\text{beam splitter}}=\left[ 
\begin{tabular}{c|c}
$0$ & $0$ \\ \hline
$0$ & $T$%
\end{tabular}
\right] \equiv T.
\end{equation*}
Our aim is to describe the effective Markov model for the feedback device
sketched below where the feedback is implemented by means of a beam
splitter. Here we have a component system, called the plant, in-loop and we
assume that it is described by the transfer function $\Xi _{0}=\left[ 
\begin{tabular}{c|c}
$A_{0}$ & $-C_{0}^{\dag }S_{0}$ \\ \hline
$C_{0}$ & $S_{0}$%
\end{tabular}
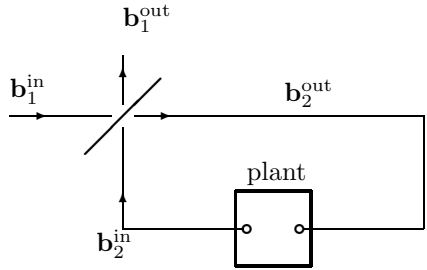
\right] $.

\begin{center}
%TCIMACRO{
%\TeXButton{figure 5}{\setlength{\unitlength}{.05cm}
%\begin{picture}(120,80)
%\label{pic5}
%
%\thicklines
%\put(30,40){\line(1,1){20}}
%
%\thinlines
%\put(10,50){\line(1,0){27}}
%\put(10,50){\vector(1,0){10}}
%\put(40,20){\line(0,1){27}}
%\put(40,20){\vector(0,1){10}}
%\put(40,53){\line(0,1){13}}
%\put(40,53){\vector(0,1){10}}
%\put(43,50){\line(1,0){77}}
%\put(43,50){\vector(1,0){10}}
%\put(120,20){\line(0,1){30}}
%\put(40,20){\line(1,0){32}}
%\put(120,20){\line(-1,0){32}}
%
%\thicklines
%
%\put(70,10){\line(0,1){20}}
%\put(70,10){\line(1,0){20}}
%\put(90,10){\line(0,1){20}}
%\put(70,30){\line(1,0){20}}
%\put(87,20){\circle{2}}
%\put(73,20){\circle{2}}
%\put(73,33){plant}
%
%
%
%\put(10,55){${\bf b}_{1}^{\rm in}$}
%\put(40,74){${\bf b}_{1}^{\rm out}$}
%\put(83,54){${\bf b}_{2}^{\rm out}$}
%
%\put(33,14){${\bf b}_{2}^{\rm in}$}
%
%\end{picture}
%}}%
%BeginExpansion
\setlength{\unitlength}{.05cm}
\begin{picture}(120,80)
\label{pic5}

\thicklines
\put(30,40){\line(1,1){20}}

\thinlines
\put(10,50){\line(1,0){27}}
\put(10,50){\vector(1,0){10}}
\put(40,20){\line(0,1){27}}
\put(40,20){\vector(0,1){10}}
\put(40,53){\line(0,1){13}}
\put(40,53){\vector(0,1){10}}
\put(43,50){\line(1,0){77}}
\put(43,50){\vector(1,0){10}}
\put(120,20){\line(0,1){30}}
\put(40,20){\line(1,0){32}}
\put(120,20){\line(-1,0){32}}

\thicklines

\put(70,10){\line(0,1){20}}
\put(70,10){\line(1,0){20}}
\put(90,10){\line(0,1){20}}
\put(70,30){\line(1,0){20}}
\put(87,20){\circle{2}}
\put(73,20){\circle{2}}
\put(73,33){plant}

\put(10,55){${\bf b}_{1}^{\rm in}$}
\put(40,74){${\bf b}_{1}^{\rm out}$}
\put(83,54){${\bf b}_{2}^{\rm out}$}

\put(33,14){${\bf b}_{2}^{\rm in}$}

\end{picture}
%
%EndExpansion

Figure 5: Feedback using a beam-splitter.
\end{center}

It is more convenient to view this as the network sketched below.

\begin{center}
%TCIMACRO{
%\TeXButton{figure 6}{\setlength{\unitlength}{.075cm}
%\begin{picture}(60,50)
%\label{pic6}
%\thicklines
%\put(26,10){$s_1$}
%\put(26,20){$s_2$}
%\put(26,40){$r_3$}
%\put(31,10){$r_1$}
%\put(31,20){$r_2$}
%\put(31,40){$s_3$}
%\put(20,5){\line(0,1){20}}
%\put(20,5){\line(1,0){20}}
%\put(40,5){\line(0,1){20}}
%\put(20,25){\line(1,0){20}}
%\put(20,35){\line(0,1){10}}
%\put(20,35){\line(1,0){20}}
%\put(40,35){\line(0,1){10}}
%\put(20,45){\line(1,0){20}}
%\thinlines
%\put(10,20){\line(1,0){13}}
%\put(24,20){\circle{2}}
%\put(10,40){\vector(1,0){5}}
%\put(10,10){\line(1,0){13}}
%\put(24,10){\circle{2}}
%\put(24,40){\circle{2}}
%\put(10,10){\vector(-1,0){5}}
%\put(37,20){\line(1,0){13}}
%\put(36,20){\circle{2}}
%\put(37,10){\line(1,0){13}}
%\put(36,10){\circle{2}}
%\put(36,40){\circle{2}}
%\put(50,20){\vector(-1,0){5}}
%\put(50,10){\vector(-1,0){5}}
%\put(10,20){\line(0,1){20}}
%\put(10,40){\line(1,0){13}}
%\put(50,40){\line(-1,0){13}}
%\put(50,20){\line(0,1){20}}
%\put(2,5){$ {\bf b}_{1}^{\rm out}$}
%\put(55,5){$ {\bf b}_{1}^{\rm in}$}
%\put(43,43){{\it plant}}
%\put(42,14){{\it beam splitter} }
%\end{picture}
%}}%
%BeginExpansion
\setlength{\unitlength}{.075cm}
\begin{picture}(60,50)
\label{pic6}
\thicklines
\put(26,10){$s_1$}
\put(26,20){$s_2$}
\put(26,40){$r_3$}
\put(31,10){$r_1$}
\put(31,20){$r_2$}
\put(31,40){$s_3$}
\put(20,5){\line(0,1){20}}
\put(20,5){\line(1,0){20}}
\put(40,5){\line(0,1){20}}
\put(20,25){\line(1,0){20}}
\put(20,35){\line(0,1){10}}
\put(20,35){\line(1,0){20}}
\put(40,35){\line(0,1){10}}
\put(20,45){\line(1,0){20}}
\thinlines
\put(10,20){\line(1,0){13}}
\put(24,20){\circle{2}}
\put(10,40){\vector(1,0){5}}
\put(10,10){\line(1,0){13}}
\put(24,10){\circle{2}}
\put(24,40){\circle{2}}
\put(10,10){\vector(-1,0){5}}
\put(37,20){\line(1,0){13}}
\put(36,20){\circle{2}}
\put(37,10){\line(1,0){13}}
\put(36,10){\circle{2}}
\put(36,40){\circle{2}}
\put(50,20){\vector(-1,0){5}}
\put(50,10){\vector(-1,0){5}}
\put(10,20){\line(0,1){20}}
\put(10,40){\line(1,0){13}}
\put(50,40){\line(-1,0){13}}
\put(50,20){\line(0,1){20}}
\put(2,5){$ {\bf b}_{1}^{\rm out}$}
\put(55,5){$ {\bf b}_{1}^{\rm in}$}
\put(43,43){{\it plant}}
\put(42,14){{\it beam splitter} }
\end{picture}
%
%EndExpansion

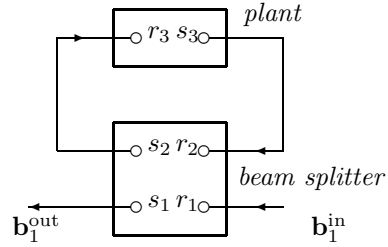
Figure 6: Network representation.
\end{center}

Here we have the pair of internal edges $\left( s_{2},r_{3}\right) $ and $%
\left( s_{3},r_{2}\right) $. The transfer function for the network is 
\begin{equation*}
\Xi _{\text{unconn.}}=\left[ 
\begin{tabular}{l|lll}
$A_{0}$ & 0 & 0 & $-C_{0}^{\dag }S_{0}$ \\ \hline
0 & $T_{11}$ & $T_{12}$ & 0 \\ 
0 & $T_{21}$ & $T_{22}$ & 0 \\ 
$C_{0}$ & 0 & 0 & $S_{0}$%
\end{tabular}
\right]
\end{equation*}
with respect to the labels $\left( 0,s_{1},s_{2},s_{3}\right) $ for the rows
and $\left( 0,r_{1},r_{2},r_{3}\right) $ for the columns. This time the
external fields are $\mathbf{b}_{\mathsf{e}}^{\text{in}}=\mathbf{b}_{1}^{%
\text{in}}$, $\mathbf{b}_{\mathsf{e}}^{\text{out}}=\mathbf{b}_{1}^{\text{out}%
}\equiv T_{11}\mathbf{b}_{1}^{\text{in}}+T_{12}\mathbf{b}_{2}^{\text{in}}$
while the (matched) internal fields are 
\begin{equation*}
\mathbf{b}_{\mathsf{i}}^{\text{in}}=\left( 
\begin{array}{c}
\mathbf{b}_{2}^{\text{in}} \\ 
\mathbf{b}_{3}^{\text{in}}
\end{array}
\right) ,\quad \mathbf{b}_{\mathsf{i}}^{\text{out}}=\left( 
\begin{array}{c}
\mathbf{b}_{2}^{\text{out}} \\ 
\mathbf{b}_{3}^{\text{out}}
\end{array}
\right) \equiv \left( 
\begin{array}{c}
T_{21}\mathbf{b}_{1}^{\text{in}}+T_{22}\mathbf{b}_{2}^{\text{in}} \\ 
S_{0}\mathbf{b}_{3}^{\text{in}}+L_{0}
\end{array}
\right) .
\end{equation*}
That is 
\begin{equation*}
\begin{tabular}{ll}
$S_{\mathsf{ii}}=\left( 
\begin{array}{cc}
T_{22} & 0 \\ 
0 & S_{0}
\end{array}
\right) ,$ & $S_{\mathsf{ie}}=\left( 
\begin{array}{c}
T_{21} \\ 
0
\end{array}
\right) ,$ \\ 
$S_{\mathsf{ei}}=\left( T_{12},0\right) ,$ & $S_{\mathsf{ee}}=T_{11},$ \\ 
$L_{\mathsf{i}}=\left( 
\begin{array}{c}
L_{0} \\ 
0
\end{array}
\right) ,$ & $L_{\mathsf{i}}=0,\quad \eta =\left( 
\begin{array}{cc}
0 & 1 \\ 
1 & 0
\end{array}
\right) .$%
\end{tabular}
\end{equation*}

Substituting into our reduction formula we obtain 
\begin{eqnarray*}
S &=&T_{11}+ 
\begin{array}{c}
(T_{12},0)
\end{array}
\left( 
\begin{array}{cc}
-T_{22} & 1 \\ 
1 & -S_{0}
\end{array}
\right) ^{-1}\left( 
\begin{array}{c}
T_{21} \\ 
0
\end{array}
\right) \\
&\equiv &T_{11}+T_{12}\left( 1-S_{0}T_{22}\right) ^{-1}S_{0}T_{21}, \\
C &=& 
\begin{array}{c}
(T_{12},0)
\end{array}
\left( 
\begin{array}{cc}
-T_{22} & 1 \\ 
1 & -S_{0}
\end{array}
\right) ^{-1}\left( 
\begin{array}{c}
0 \\ 
C_{0}
\end{array}
\right) \\
&\equiv &T_{12}\left( 1-S_{0}T_{22}\right) ^{-1}C_{0}, \\
\Omega &=&\Omega _{0}+\func{Im} 
\begin{array}{c}
(0,L_{0}^{\dag })
\end{array}
\left( 
\begin{array}{cc}
-T_{22} & 1 \\ 
1 & -S_{0}
\end{array}
\right) ^{-1}\left( 
\begin{array}{c}
0 \\ 
L_{0}
\end{array}
\right) \\
&\equiv &\Omega _{0}+\func{Im}C_{0}^{\dag }\left( 1-S_{0}T_{22}\right)
^{-1}C_{0}.
\end{eqnarray*}
and so, when the connections are made, the transfer fmatrix function is 
\begin{equation*}
\Xi _{\text{conn.}}=\left[ 
\begin{tabular}{l|l}
$A_{0}-C_{0}^{\dag }S_{0}T_{22}C_{0}$ & $-C_{0}^{\dag
}S_{0}T_{21}-C_{0}^{\dag }S_{0}\left( 1-S_{0}T_{22}\right) ^{-1}T_{22}$ \\ 
\hline
$T_{12}\left( 1-S_{0}T_{22}\right) ^{-1}C_{0}$ & $T_{11}+T_{12}\left(
1-S_{0}T_{22}\right) ^{-1}S_{0}T_{21}$%
\end{tabular}
\right] ,
\end{equation*}

Note that $S=\varphi _{T}^{2\rightarrow 1}\left( S_{0}\right) $ where $%
\varphi _{T}^{2\rightarrow 1}\left( z\right) =T_{11}+T_{12}\beta \left(
z^{-1}-T_{22}\right) T_{21}$ is the M\"{o}bius transformation in the complex
plane associated with $T$.

If we further set $T=\left( 
\begin{array}{cc}
\alpha & \beta \\ 
\mu & \nu
\end{array}
\right) $, and $x+iy=S_{0}\nu $, then 
\begin{gather*}
C^{\dag }C=\left| \frac{\beta }{1-S_{0}\nu }\right| ^{2}C_{0}^{\dag }C_{0}=%
\frac{1-|\nu |^{2}}{|1-S_{0}\nu |^{2}}C_{0}^{\dag }C_{0}\equiv \frac{%
1-x^{2}-y^{2}}{\left( 1-x\right) ^{2}+y^{2}}C_{0}^{\dag }C_{0}, \\
\func{Im}C_{0}^{\dag }\left( 1-S_{0}\nu \right) ^{-1}C_{0}=\func{Im}\left\{ 
\frac{1}{1-x-iy}\right\} C_{0}^{\dag }C_{0}=\frac{y}{\left( 1-x\right)
^{2}+y^{2}}C_{0}^{\dag }C_{0}.
\end{gather*}
In particular, if we take a single oscillator in-loop with $S_{0}=e^{i\phi
_{0}}$, then we obtain $S\equiv e^{i\phi }$ and the phase is determined by
the M\"{o}bius transformation. If we further have $L_{0}=\sqrt{\gamma _{0}}a$%
, $H_{0}=\omega _{0}a^{\dag }a$, we find that $L\equiv e^{i\delta }\sqrt{%
\gamma }a$ and $H=\omega a^{\dag }a$ where 
\begin{equation*}
\gamma =\frac{1-x^{2}-y^{2}}{\left( 1-x\right) ^{2}+y^{2}}\gamma _{0},\quad
\omega =\frac{y}{\left( 1-x\right) ^{2}+y^{2}}\omega _{0},
\end{equation*}
and $\delta $ is a real phase. In the specific case $T=\left( 
\begin{array}{cc}
\alpha & \beta \\ 
\beta & -\alpha
\end{array}
\right) $ with $S_{0}=1,\omega _{0}=0$\ considered by Yanagisawa and Kimura 
\cite{YK1}, we have $x=-\alpha $ and $y=0$, therefore we find 
\begin{equation*}
\gamma =\frac{1-\alpha }{1+\alpha }\gamma _{0},\quad \omega =0
\end{equation*}
which agrees with their findings.

\bigskip

An alternative computation of $\Xi $ \ is given by the following argument.
We consider the input-output relations 
\begin{equation*}
\mathbf{\hat{b}}_{i}^{\text{out}}=\sum_{j=1,2}T_{ij}\mathbf{\hat{b}}_{j}^{%
\text{in}},\quad \mathbf{\hat{b}}_{2}^{\text{in}}=\Xi _{0}\mathbf{\hat{b}}%
_{1}^{\text{out}}+\xi _{0}\mathbf{a}_{0},
\end{equation*}
and eliminating $\mathbf{\hat{b}}_{2}^{\text{out}}\equiv \left( 1-T_{22}\Xi
\right) ^{-1}\left[ T_{21}\mathbf{\hat{b}}_{1}^{\text{in}}+T_{22}\xi _{0}%
\mathbf{a}_{0}\right] $ yields 
\begin{equation*}
\mathbf{\hat{b}}_{1}^{\text{out}}=\left[ T_{11}+T_{12}\Xi _{0}\left(
1-T_{22}\Xi _{0}\right) ^{-1}T_{21}\right] \mathbf{\hat{b}}_{1}^{\text{in}%
}+T_{21}\left( 1-\Xi _{0}T_{22}\right) ^{-1}\xi _{0}\mathbf{a}_{0}.
\end{equation*}
That is 
\begin{equation*}
\Xi =T_{11}+T_{12}\left( \Xi _{0}^{-1}-T_{22}\right) ^{-1}T_{21}=\varphi
_{T}^{2\rightarrow 1}\left( \Xi _{0}\right) .
\end{equation*}
We remark that if $T_{12}$ and $T_{21}$ are invertible, then we may invert
the M\"{o}bius transformation to get 
\begin{equation*}
\Xi _{0}^{-1}=T_{22}+T_{21}\frac{1}{\Xi -T_{11}}T_{12}.
\end{equation*}

To illustrate with a cavity mode in-loop, we take the beam splitter matrix
to be $T=\left( 
\begin{array}{cc}
\alpha & \beta \\ 
\beta & -\alpha
\end{array}
\right) $ with $\alpha ^{2}+\beta ^{2}=1$, and the transfer function $\Xi
_{0}\left( s\right) =\frac{s+i\omega -\gamma /2}{s+i\omega +\gamma /2}$,
then we find 
\begin{equation*}
\Xi =\frac{\alpha +\Xi _{0}}{1+\alpha \Xi _{0}}=\frac{s+i\omega -\frac{%
1-\alpha }{1+\alpha }\frac{\gamma }{2}}{s+i\omega +\frac{1-\alpha }{1+\alpha 
}\frac{\gamma }{2}}.
\end{equation*}

\section{The Redheffer Star Product}

An important feedback arrangement is shown in the figure below.

\begin{center}
%TCIMACRO{
%\TeXButton{figure 7}{\setlength{\unitlength}{.1cm}
%\begin{picture}(60,65)
%\label{pic7}
%\thicklines
%\put(20,10){\line(0,1){20}}
%\put(20,10){\line(1,0){20}}
%\put(40,10){\line(0,1){20}}
%\put(20,30){\line(1,0){20}}
%\put(20,40){\line(0,1){20}}
%\put(20,40){\line(1,0){20}}
%\put(40,40){\line(0,1){20}}
%\put(20,60){\line(1,0){20}}
%\thinlines
%\put(24,14){\circle{2}}
%\put(24,26){\circle{2}}
%\put(36,14){\circle{2}}
%\put(36,26){\circle{2}}
%\put(24,44){\circle{2}}
%\put(24,56){\circle{2}}
%\put(36,44){\circle{2}}
%\put(36,56){\circle{2}}
%\put(23,14){\vector(-1,0){12}}
%\put(49,14){\vector(-1,0){12}}
%\put(11,56){\vector(1,0){12}}
%\put(37,56){\vector(1,0){12}}
%\put(23,26){\line(-1,0){8}}
%\put(15,26){\vector(0,1){18}}
%\put(23,44){\line(-1,0){8}}
%\put(37,26){\line(1,0){8}}
%\put(45,44){\vector(0,-1){18}}
%\put(37,44){\line(1,0){8}}
%\put(28,62){$A$}
%\put(0,12){${\bf b}^{\rm out}_4$}
%\put(-5,35){$ {\bf b}^{\rm out}_3  ={\bf b}^{\rm in}_2$}
%\put(0,56){${\bf b}^{\rm in}_1$}
%\put(28,4){$B$}
%\put(50,12){${\bf b}^{\rm in}_4$}
%\put(50,35){${\bf b}^{\rm out}_2 = {\bf b}^{\rm in}_3$}
%\put(50,56){${\bf b}^{\rm out}_1$}
%\put(26,14){$s_4$}
%\put(26,26){$s_3$}
%\put(26,44){$r_2$}
%\put(26,56){$r_1$}
%\put(31,14){$r_4$}
%\put(31,26){$r_3$}
%\put(31,44){$s_2$}
%\put(31,56){$s_1$}
%\end{picture}%
%}}%
%BeginExpansion
\setlength{\unitlength}{.1cm}
\begin{picture}(60,65)
\label{pic7}
\thicklines
\put(20,10){\line(0,1){20}}
\put(20,10){\line(1,0){20}}
\put(40,10){\line(0,1){20}}
\put(20,30){\line(1,0){20}}
\put(20,40){\line(0,1){20}}
\put(20,40){\line(1,0){20}}
\put(40,40){\line(0,1){20}}
\put(20,60){\line(1,0){20}}
\thinlines
\put(24,14){\circle{2}}
\put(24,26){\circle{2}}
\put(36,14){\circle{2}}
\put(36,26){\circle{2}}
\put(24,44){\circle{2}}
\put(24,56){\circle{2}}
\put(36,44){\circle{2}}
\put(36,56){\circle{2}}
\put(23,14){\vector(-1,0){12}}
\put(49,14){\vector(-1,0){12}}
\put(11,56){\vector(1,0){12}}
\put(37,56){\vector(1,0){12}}
\put(23,26){\line(-1,0){8}}
\put(15,26){\vector(0,1){18}}
\put(23,44){\line(-1,0){8}}
\put(37,26){\line(1,0){8}}
\put(45,44){\vector(0,-1){18}}
\put(37,44){\line(1,0){8}}
\put(28,62){$A$}
\put(0,12){${\bf b}^{\rm out}_4$}
\put(-5,35){$ {\bf b}^{\rm out}_3  ={\bf b}^{\rm in}_2$}
\put(0,56){${\bf b}^{\rm in}_1$}
\put(28,4){$B$}
\put(50,12){${\bf b}^{\rm in}_4$}
\put(50,35){${\bf b}^{\rm out}_2 = {\bf b}^{\rm in}_3$}
\put(50,56){${\bf b}^{\rm out}_1$}
\put(26,14){$s_4$}
\put(26,26){$s_3$}
\put(26,44){$r_2$}
\put(26,56){$r_1$}
\put(31,14){$r_4$}
\put(31,26){$r_3$}
\put(31,44){$s_2$}
\put(31,56){$s_1$}
\end{picture}%
%
%EndExpansion

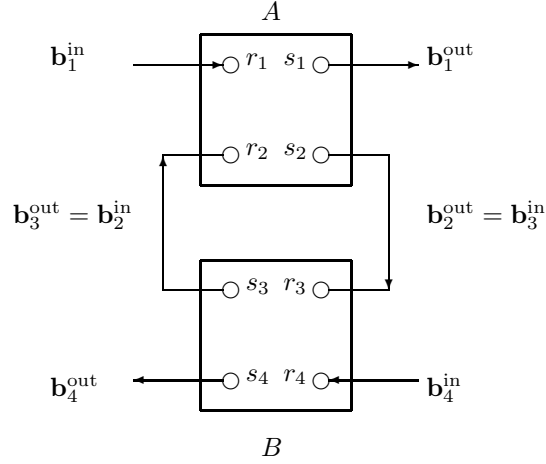
Figure 7: Composite System
\end{center}

We shall now derive the matrices for this system taking component $A$ to be
described $\left( 
\begin{array}{cc}
S_{11}^{A} & S_{12}^{A} \\ 
S_{21}^{A} & S_{22}^{A}
\end{array}
\right) ,$ $\left( 
\begin{array}{c}
C_{1}^{A} \\ 
C_{2}^{A}
\end{array}
\right) ,$ $\Omega _{A}$ and $B$ by $\left( 
\begin{array}{cc}
S_{33}^{B} & S_{34}^{B} \\ 
S_{43}^{B} & S_{44}^{B}
\end{array}
\right) ,$ $\left( 
\begin{array}{c}
C_{3}^{B} \\ 
C_{4}^{B}
\end{array}
\right) ,$ $\Omega _{B}$. The operators of systems $A$ are asumed to commute
with those of $B$. We have two internal channels to eliminate which we can
do in sequence, or simulataneously. We shall do the latter. here we have 
\begin{eqnarray*}
\mathsf{S}_{\mathtt{ee}} &=&\left( 
\begin{array}{cc}
S_{11}^{A} & 0 \\ 
0 & S_{44}
\end{array}
\right) ,\mathsf{S}_{\mathtt{ei}}=\left( 
\begin{array}{cc}
S_{12}^{A} & 0 \\ 
0 & S_{43}^{B}
\end{array}
\right) \\
\mathsf{S}_{\mathtt{ie}} &=&\left( 
\begin{array}{cc}
S_{21}^{A} & 0 \\ 
0 & S_{34}^{B}
\end{array}
\right) ,\mathsf{S}_{\mathtt{ii}}=\left( 
\begin{array}{cc}
S_{22}^{A} & 0 \\ 
0 & S_{33}^{B}
\end{array}
\right)
\end{eqnarray*}
and 
\begin{equation*}
\mathsf{L}_{\mathtt{e}}=\left( 
\begin{array}{c}
L_{1}^{A} \\ 
L_{4}^{B}
\end{array}
\right) ,\quad \mathsf{L}_{\mathtt{i}}=\left( 
\begin{array}{c}
L_{2}^{A} \\ 
L_{3}^{B}
\end{array}
\right) ,\quad \eta =\left( 
\begin{array}{cc}
0 & 1 \\ 
1 & 0
\end{array}
\right) .
\end{equation*}
The parameters are therefore 
\begin{eqnarray*}
S_{\star } &=&\left( 
\begin{array}{cc}
S_{11}^{A} & 0 \\ 
0 & S_{44}^{B}
\end{array}
\right) +\left( 
\begin{array}{cc}
S_{12}^{A} & 0 \\ 
0 & S_{43}^{B}
\end{array}
\right) \left( 
\begin{array}{cc}
-S_{22}^{A} & 1 \\ 
1 & -S_{33}^{B}
\end{array}
\right) ^{-1}\left( 
\begin{array}{cc}
S_{21}^{A} & 0 \\ 
0 & S_{34}^{B}
\end{array}
\right) \\
&=&\left( 
\begin{array}{cc}
S_{11}^{A}+S_{12}^{A}S_{33}^{B}\left( 1-S_{22}^{A}S_{33}^{B}\right)
^{-1}S_{21}^{A} & S_{12}^{A}\left( 1-S_{22}^{A}S_{33}^{B}\right) ^{-1}S_{34}
\\ 
S_{43}^{B}\left( 1-S_{22}^{A}S_{33}^{B}\right) ^{-1}S_{21}^{A} & 
S_{44}+S_{43}^{B}\left( 1-S_{22}^{A}S_{33}^{B}\right) ^{-1}S_{22}^{A}S_{34}
\end{array}
\right) , \\
C_{\star } &=&\left( 
\begin{array}{c}
C_{1}^{A} \\ 
C_{4}^{B}
\end{array}
\right) +\left( 
\begin{array}{cc}
S_{12}^{A} & 0 \\ 
0 & S_{43}^{B}
\end{array}
\right) \left( 
\begin{array}{cc}
-S_{22}^{A} & 1 \\ 
1 & -S_{33}^{B}
\end{array}
\right) ^{-1}\left( 
\begin{array}{c}
C_{2}^{A} \\ 
C_{3}^{B}
\end{array}
\right) \\
&=&\left( 
\begin{array}{c}
C_{1}^{A}+S_{12}^{A}S_{33}^{B}\left( 1-S_{22}^{A}S_{33}^{B}\right)
^{-1}C_{2}^{A}+S_{12}^{A}\left( 1-S_{22}^{A}S_{33}^{B}\right) ^{-1}C_{3}^{B}
\\ 
C_{4}^{B}+S_{43}^{B}\left( 1-S_{22}^{A}S_{33}^{B}\right)
^{-1}C_{2}^{A}+S_{43}^{B}S_{22}^{A}\left( 1-S_{22}^{A}S_{33}^{B}\right)
^{-1}C_{3}^{B}
\end{array}
\right) ,
\end{eqnarray*}
\begin{equation*}
\begin{array}{l}
\Omega _{\star }=\Omega _{A}+\Omega _{B}+\func{Im}\left\{ C_{3}^{B\dag
}\left( 1-S_{33}^{B}S_{22}^{A}\right) ^{-1}C_{3}^{B}+C_{3}^{B\dag }\left(
1-S_{33}^{B}S_{22}^{A}\right) ^{-1}S_{33}^{A}C_{2}^{A}\right. \\ 
+C_{2}^{A\dag }\left( 1-S_{22}^{A}S_{33}^{B}\right)
^{-1}S_{22}^{A}C_{3}^{B}+C_{2}^{A\dag }\left( 1-S_{22}^{A}S_{33}^{B}\right)
^{-1}C_{2}^{A} \\ 
+C_{1}^{A\dag }S_{12}^{A}\left( 1-S_{33}^{B}S_{22}^{A}\right)
^{-1}C_{3}^{B}+C_{1}^{A\dag }S_{12}^{A}\left( 1-S_{33}^{B}S_{22}^{A}\right)
^{-1}S_{33}^{B}C_{2}^{B} \\ 
\left. +C_{4}^{B\dag }S_{43}^{B}\left( 1-S_{22}^{A}S_{33}^{B}\right)
^{-1}S_{22}^{A}C_{3}^{B}+C_{4}^{B\dag }S_{43}^{B}\left(
1-S_{22}^{A}S_{33}^{B}\right) ^{-1}C_{2}^{A}\right\} .
\end{array}
\end{equation*}

\section{Appendix (Stratonovich to It\={o} Conversion)}

It is convenient to introduce integrated fields 
\begin{equation*}
B_{i}\left( t\right) \equiv \int_{0}^{t}b_{i}\left( s\right) ds,B_{i}^{\dag
}\left( t\right) \equiv \int_{0}^{t}b_{i}^{\dag }\left( s\right) ds,\Lambda
_{ij}\left( t\right) \equiv \int_{0}^{t}b_{i}^{\dag }\left( s\right)
b_{j}\left( s\right) ds.
\end{equation*}
$B_{i}\left( t\right) $ and $B_{i}^{\dag }\left( t\right) $ are called the
annihilation and creation process, respectively, for the $i$th field and
collectivey are referred to as a quantum Wiener process. $\Lambda
_{ij}\left( t\right) $ is called the gauge process or scattering process
from the $j$th field to the $i$th field. A noncommutative version of the Ito
theory of stochastic integration with respect to these processes can be
built up. The quantum It\={o} table giving the product of infinitesimal
increments of these process is 
\begin{equation*}
\begin{tabular}{l|llll}
$\times $ & $dB_{k}$ & $d\Lambda _{kl}$ & $dB_{l}^{\dag }$ & $dt$ \\ \hline
$dB_{i}$ & 0 & $\delta _{ik}dB_{l}$ & $\delta _{il}dt$ & 0 \\ 
$d\Lambda _{ij}$ & 0 & $\delta _{jk}d\Lambda _{il}$ & $\delta
_{jl}dB_{i}^{\dag }$ & 0 \\ 
$dB_{j}^{\dag }$ & 0 & 0 & 0 & 0 \\ 
$dt$ & 0 & 0 & 0 & 0
\end{tabular}
.
\end{equation*}
The Ito equation for the unitary process is then $dV=\left( dG\right) V$
where 
\begin{equation*}
dG=\sum_{i,j=1}^{n}(S_{ij}-\delta _{ij})d\Lambda
_{ij}+\sum_{i=1}^{n}L_{i}dB_{i}^{\dag }-\sum_{j=1}^{n}L_{i}^{\dag
}S_{ij}B_{j}-\left( \frac{1}{2}\sum_{i=1}^{n}L_{i}^{\dag }L_{i}-iH\right) dt.
\end{equation*}
The Stratonovich form is $dV=-i\left( dE\right) \circ V$ where 
\begin{equation*}
dE=\sum_{i,j=1}^{n}E_{ij}d\Lambda _{ij}+\sum_{i=1}^{n}F_{i}dB_{i}^{\dag
}+\sum_{j=1}^{n}F_{j}^{\dag }dB_{j}\left( t\right) +Kdt.
\end{equation*}
and we define the Stratonovich differential to be $\left( dX\right) \circ
Y=\left( dX\right) Y+\frac{1}{2}\left( dX\right) \left( dY\right) $ with the
last term computed using the It\={o} table. We have the consistency
condition $dV=\left( dG\right) V\equiv -i\left( dE\right) V-\frac{i}{2}%
\left( dE\right) \left( dG\right) V$ or 
\begin{equation*}
dG=-idE-\frac{i}{2}\left( dE\right) \left( dG\right) ,
\end{equation*}
and using the table we see that 
\begin{eqnarray*}
S-1 &\equiv &-iE-\frac{i}{2}E\left( S-1\right) \\
L &=&-iF-\frac{i}{2}EL \\
-\frac{1}{2}L^{\dag }L-iH &=&-iK-\frac{i}{2}F^{\dag }L
\end{eqnarray*}
which can be solved to give the relations $\left( \ref{Strat-Ito}\right) $.

\end{document}